\documentclass[aps,prl,twocolumn,twoside,showpacs]{revtex4}
\usepackage{graphicx,amssymb,amsmath,epsf,bm}
\epsfclipon

\newcommand{\rd}{\mathrm{d}}
\newcommand{\expct}[1]{\langle #1 \rangle}

\newcommand{\prt}[2]{\frac{\partial #1}{\partial #2}}

\newcommand{\unit}[1]{\,\mathrm{#1}}

\begin{document}

\title{Universal Fluctuations of Growing Interfaces: Evidence in Turbulent Liquid Crystals}

\author{Kazumasa A. Takeuchi}
 \email[]{kazumasa@daisy.phys.s.u-tokyo.ac.jp}
\author{Masaki Sano}
%\affiliation{Service de Physique de l'\'Etat Condens\'e,~CEA -- Saclay,~91191~Gif-sur-Yvette,~France}%
\affiliation{Department of Physics,\,The University of Tokyo,\,7-3-1 Hongo,\,Bunkyo-ku,\,Tokyo 113-0033,\,Japan}%

\date{\today}

\begin{abstract}
We investigate
 growing interfaces of topological-defect turbulence
 in the electroconvection of nematic liquid crystals.
The interfaces exhibit self-affine roughening
 characterized by both spatial and temporal scaling laws
 of the Kardar-Parisi-Zhang theory in 1+1 dimensions.
Moreover, we reveal that the distribution
 and the two-point correlation of the interface fluctuations
 are universal ones
 governed by the largest eigenvalue of random matrices.
%Moreover, we reveal universal distribution
% and correlation functions of the interface fluctuations,
% which are governed by the largest eigenvalue of random matrices.
This provides quantitative experimental evidence
 of the universality prescribing
 detailed information of scale-invariant fluctuations.

\end{abstract}


\maketitle

Growth phenomena have been a subject of extensive studies
 in physics and beyond, because of the ubiquity in nature
 and of their importance in both engineering and fundamental science.
Over decades, physicists have found that
 growth phenomena due to local processes
 typically lead to the formation of rough self-affine interfaces,
 as exemplified in paper wetting, burning fronts,
 bacterial colonies, and material morphology,
 to name but a few, and also in various numerical models
 \cite{Barabasi_Stanley-Book1995}.
Being obviously irreversible, local growth processes provide
 a challenging situation toward understanding the scale invariance
 and the consequent universality out of equilibrium.

The roughness of interfaces is often quantified by their width $w(l,t)$
 defined as the standard deviation of the interface height $h(x,t)$
 over a length scale $l$ at time $t$.
% i.e., $w(l,t) \equiv \expct{\sqrt{\expct{[h(x,t)-\expct{h}_l]^2}_l}}$,
% where $\expct{\cdots}_l$ and $\expct{\cdots}$ denote
% the average over a segment of length $l$, and all over the interface
% and ensembles, respectively.
The self-affinity of interfaces then implies
 the following Family-Vicsek scaling \cite{Family_Vicsek-JPhysA1985}:
\vspace{-7pt}
\begin{equation}
 w(l,t) \sim t^{\beta} F(lt^{-1/z}) \sim \begin{cases} l^\alpha & \text{for $l \ll l_*$}, \\ t^{\beta} & \text{for $l \gg l_*$}, \end{cases}  \label{eq:FVScaling}
\end{equation}
 with two characteristic exponents $\alpha$ and $\beta$,
% called the roughness and growth exponents, respectively,
 the dynamic exponent $z \equiv \alpha/\beta$,
 and a crossover length scale $l_* \sim t^{1/z}$.

The simplest
% continuum
 theory to describe such local growth processes
 was proposed by Kardar, Parisi, and Zhang (KPZ) \cite{Kardar_etal-PRL1986}
 on the basis of the coarse-grained stochastic equation
\begin{equation}
 \prt{}{t}h(x,t) = v_0 + \nu \nabla^2 h + \frac{\lambda}{2}(\nabla h)^2 + \xi(x,t)  \label{eq:KPZ}
\end{equation}
 with
% white noise
 $\expct{\xi(x,t)} = 0$ and
 $\expct{\xi(x,t)\xi(x',t')} = D \delta(x-x')\delta(t-t')$.
%Exact values of the exponents are accessible in 1+1 dimensions,
For 1+1 dimensions, the renormalization group approach provides
 exact values of the exponents
 at $\alpha^\mathrm{KPZ} = 1/2$ and $\beta^\mathrm{KPZ} = 1/3$
 \cite{Kardar_etal-PRL1986,Barabasi_Stanley-Book1995},
 which are universal as widely confirmed in numerical models
 \cite{Barabasi_Stanley-Book1995}.
Moreover, the (1+1)-dimensional KPZ class attracts growing interest
 thanks to rigorous work on the asymptotic form
% has a solvable member which allows to calculate the asymptotic distribution
 of the fluctuations in solvable models
 \cite{Spohn-PhysA2006,Johansson-CommunMathPhys2000,Prahofer_Spohn-2000}.
% which is also expected to be shared universally.
This opens up a new aspect in the study of scale-invariant phenomena
 toward the universality beyond the scaling laws.
%It is elegantly described by
%% the largest eigenvalue of
% random matrices
%% whose ensemble is determined by the geometry of the interfaces,
% and expected to be shared universally in the KPZ class.

In contrast with such remarkable progress in theory,
 the situation in experiments has been quite different.
A considerable number of experiments have been performed
 on various growth processes \cite{Barabasi_Stanley-Book1995}
 and confirmed the ubiquity of rough interfaces.
% that rough interfaces are indeed ubiquitous.
Concerning the universality, however,
 experimentally measured values of the exponents
 are widely diverse and mostly far from the KPZ values
 for both $\alpha$ and $\beta$ \cite{Barabasi_Stanley-Book1995}.
To our knowledge, only two experiments among dozens
 directly found the KPZ exponents:
% in bacterial colonies of mutant \textit{Bacillus subtilis}
 in colonies of mutant bacteria
 \cite{Wakita_etal-JPSJ1997}
 and in slow combustion of paper
 \cite{Maunuksela_etal-PRL1997},
 apart from few other experiments showing
 indirect indications \cite{Degawa_etal-PRL2006,DirectedPolymer}.
One of the main difficulties shared by most experiments,
 including the above two,
% in favor of the KPZ behavior,
 is that one needs to repeat a large number of experiments
 in the same controlled conditions to accumulate sufficient statistics.
In this Letter, studying growing interfaces of turbulent liquid crystals,
 we overcome this difficulty and report
% clear experimental results
% of critical comparisons with the wealth of theoretical predictions
% on the KPZ scaling and beyond.
% on the KPZ class.
 clear experimental evidence of not only the universal scaling laws
 but also the universal fluctuations of the KPZ class
 through critical comparisons with the wealth of theoretical predictions.

The electroconvection
% of nematic liquid crystals
 occurs when an external voltage is applied to a thin layer
 of nematic liquid crystal,
 triggering the Carr-Helfrich instability \cite{deGennes_Prost-Book1993}.
We focus on interfaces between two topologically different turbulent states
 called the dynamic scattering modes 1 and 2 (DSM1 and DSM2),
 which are observed with sufficiently large voltages.
The essential difference between them lies in the density
 of topological defects called the disclinations.
Upon applying a voltage, we first observe the DSM1 state
 with practically no defects in the director field,
 which lasts until a disclination is finally created
 owing to the breakdown of surface anchoring
 \cite{Fazio_Komitov-EurophysLett1999}.
% typically on the boundary or from impurities of the cell.
This forms a DSM2 cluster composed of a large quantity of disclinations,
 which are constantly elongated, split, and transported
 by fluctuating turbulent flow around.
While DSM2 may coexist with DSM1 in a regime of spatiotemporal intermittency
 \cite{Takeuchi_etal},
 for larger voltages we observe growing DSM2 clusters
 driven by the above-mentioned stochastic local contamination processes.

Our experimental setup consists of a quasi-two-dimensional sample cell,
 an optical microscope, a thermocontroller,
 and an ultraviolet pulse laser
%, while detailed descriptions
% of the setup were given in Ref.\ \cite{Takeuchi_etal}.
 (see Ref.\ \cite{Takeuchi_etal} for detailed descriptions).
The cell is made of
 two parallel glass plates with transparent electrodes,
 which are spaced by a polyester film of thickness $12\unit{\mu{}m}$
 enclosing a region of $16\unit{mm} \times 16\unit{mm}$
 for the convection.
We chose here the homeotropic alignment of liquid crystals
 in order to work with isotropic DSM2 growth, which is realized by coating
 $N$,$N$-dimethyl-$N$-octadecyl-3-aminopropyltrimethoxysilyl chloride
 uniformly on the electrodes using a spin coater.
The cell is then filled with $N$-(4-methoxybenzylidene)-4-butylaniline
 doped with 0.01 wt.\% of tetra-$n$-butylammonium bromide.
The cutoff frequency of the conductive regime \cite{deGennes_Prost-Book1993}
 is $850 \pm 50\unit{Hz}$.
The cell is maintained at a constant temperature $25.0\unit{^\circ{}C}$
 with typical fluctuations in the order of $10^{-3} \unit{K}$.
The convection is observed through the transmitted light
 from light-emitting diodes and recorded by a CCD camera.

\begin{figure}[t]
 \includegraphics[width=0.85\hsize,clip]{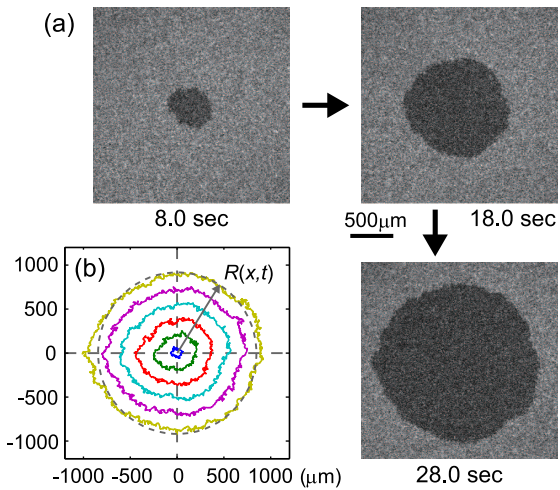}
 \caption{(Color online) Growing DSM2 cluster. (a) Images. Indicated below is the elapsed time after the emission of laser pulses. (b) Snapshots of the interfaces taken every $5\unit{s}$ in the range $2\unit{s} \leq t \leq 27\unit{s}$. The gray dashed circle shows the mean radius of all the droplets at $t=27\unit{s}$. The coordinate $x$ at this time is defined along this circle.\vspace{-5pt}}
 \label{fig1}
\end{figure}%

For each run
 we apply a voltage of $26\unit{V}$ at $250\unit{Hz}$,
% which is roughly 30\% of the cutoff frequency $850 \pm 50\unit{Hz}$
% for the convective regime of the electroconvection
% \cite{deGennes_Prost-Book1993}.
 which is sufficiently larger
 than the DSM1-DSM2 threshold at $20.7\unit{V}$.
After waiting a few seconds, we shoot into the cell
 two successive laser pulses
 of wavelength $355\unit{nm}$ and energy $6 \unit{nJ}$
% focused by a $\times 4$ objective lens,
 to trigger a DSM2 nucleus \cite{Takeuchi_etal}.
Figure \ref{fig1} displays typical growth of a DSM2 cluster.
% showing that the interfacial roughness indeed grows as time $t$ elapses.
%It is recorded by a CCD camera
% until it occupies the whole field of view
% of size $3.0\unit{mm} \times 2.3\unit{mm}$.
We repeat it 563 times to characterize the growth process precisely.

We define the local radius $R(x,t)$ along the circle
 which denotes the statistically averaged shape
 of the droplets, as sketched in Fig.\ \ref{fig1}(b).
This measures the interfacial width
% $w(l,t)$ as
% the standard deviation of $R(x,t)$ over a length scale $l$, i.e.,
 $w(l,t) \equiv \expct{\sqrt{\expct{[R(x,t)-\expct{R}_l]^2}_l}}$
 and the height-difference correlation function
% $C(l,t)$ as
 $C(l,t) \equiv \expct{[R(x+l,t) - R(x,t)]^2}$,
 where $\expct{\cdots}_l$ and $\expct{\cdots}$ denote
 the average over a segment of length $l$ and all over the interface
 and ensembles, respectively.
Both $w(l,t)$ and $C(l,t)^{1/2}$ are common quantities
 for characterizing the roughness,
 for which the Family-Vicsek scaling [Eq.\ \eqref{eq:FVScaling}] is expected.

\begin{figure}[t]
 \includegraphics[width=0.9\hsize,clip]{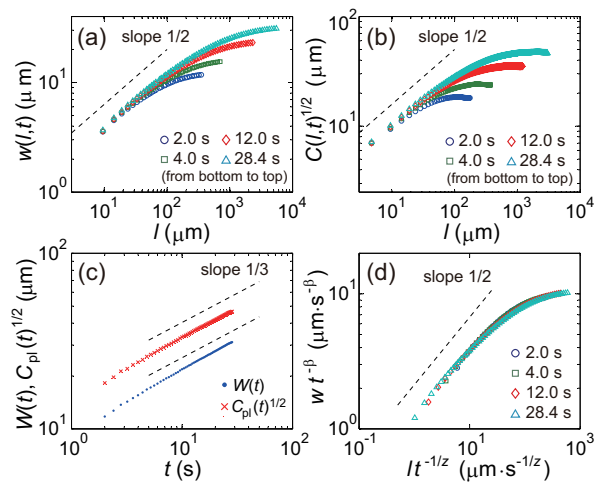}
 \caption{(Color online) Scaling of the width $w(l,t)$ and the height-difference correlation function $C(l,t)$. (a,b) Raw data of $w(l,t)$ (a) and $C(l,t)^{1/2}$ (b) at different times $t$. The length scale $l$ is varied up to $2\pi\expct{R}$ and $\pi\expct{R}$, respectively. (c) Time evolution of the overall width $W(t)$ and the plateau level $C_{\rm pl}(t)^{1/2}$ of the correlation function. (d) Collapse of the data in (a) showing the Family-Vicsek scaling [Eq.\ \eqref{eq:FVScaling}]. The dashed lines are guides for the eyes showing the KPZ scaling.\vspace{-5pt}}
 \label{fig2}
\end{figure}%

This is tested in Fig.\ \ref{fig2}.
Raw data of $w(l,t)$ and $C(l,t)^{1/2}$ measured at different times
 [Fig.\ \ref{fig2}(a,b)]
 grow algebraically for short length scales $l \ll l_*$
 and converge to constants for $l \gg l_*$
 in agreement with Eq.\ \eqref{eq:FVScaling}.
The power $\alpha$ of the algebraic regime
 measured in the last frame $t = 28.4\unit{s}$ is found to be
 $\alpha = 0.50(5)$.
Here, the number in the parentheses indicates the range of error
 in the last digit,
 which is estimated both from the uncertainty in a single fit
 and from the dependence on the fitting range.
The found value of $\alpha$ is in good agreement
 with the KPZ roughness exponent $\alpha^\mathrm{KPZ} = 1/2$.

The temporal growth of the roughness is measured by the overall width
 $W(t) \equiv \sqrt{\expct{[R(x,t)-\expct{R}]^2}}$
 and the plateau level of the correlation function, $C_{\rm pl}(t)^{1/2}$,
 defined as the mean value
 of $C(l,t)^{1/2}$ in the plateau region of Fig.\ \ref{fig2}(b).
Both quantities show a very clear power law $t^\beta$
 with $\beta = 0.336(11)$ [Fig.\ \ref{fig2}(c)]
 in remarkable agreement with the KPZ
 growth exponent $\beta^\mathrm{KPZ} = 1/3$.
Furthermore, rescaling both axes in Fig.\ \ref{fig2}(a)
% as $lt^{-1/z}$ and $w(l,t)t^{-\beta}$
 with the KPZ exponents,
 we confirm that our data of $w(l,t)$ collapse
 reasonably well onto a single curve [Fig.\ \ref{fig2}(d)].
A collapse of the same quality is obtained for $C(l,t)^{1/2}$.
We therefore safely conclude that
 the DSM2 interfacial growth belongs to the (1+1)-dimensional KPZ class.
% with respect to both $\alpha$ and $\beta$.
In passing, this rules out the logarithmic temporal scaling claimed
 by Escudero for the droplet geometry \cite{Escudero-PRL2008}.
% but favors a calculation by Singha \cite{Singha-JStatMech2005}
% which predicted the same power law as in the flat interfaces
% (with a different prefactor).

Our statistically clean data motivate us to test further predictions
 on the KPZ class
 beyond those for the scaling.
In this respect one of the most challenging benchmarks may be
 the asymptotic distribution of height fluctuations,
 calculated exactly for solvable models
 \cite{Johansson-CommunMathPhys2000,Prahofer_Spohn-2000}.
% shown by Johansson \cite{Johansson-CommunMathPhys2000} and
%% in more general contexts
% by Pr\"ahofer and Spohn \cite{Prahofer_Spohn-2000}
% (hereafter JPS).
A general expression was proposed
 by Pr\"ahofer and Spohn \cite{Prahofer_Spohn-2000},
 which reads
%\begin{equation}
 $h(t) \simeq v_\infty t + (A^2\lambda t/2)^{1/3} \chi$
%\end{equation}
 with $A \equiv D/2\nu$, the asymptotic growth rate $v_\infty$,
 and a random variable $\chi$
% which obeys the largest eigenvalue distribution of random matrices
% in the Gaussian unitary and orthogonal ensemble (GUE and GOE)
% \cite{Mehta-Book2004}
 obeying the Tracy-Widom (TW) distribution
 \cite{Tracy_Widom-CommunMathPhys},
 or the (rescaled) largest eigenvalue distribution
 of large random matrices.
The random matrices are
 from the Gaussian unitary and orthogonal ensemble (GUE and GOE)
 \cite{Mehta-Book2004}
 for curved and flat interfaces, respectively.
This implies an intriguing relation to the random matrix theory
 and requires no fitting parameter provided that the values
 of the two KPZ parameters $\lambda$ and $A$ are measured.
The prediction was tested once for flat interfaces
 in the paper combustion experiment
 \cite{Miettinen_etal-EPJB2005} with an apparent agreement.
However, the authors had to shift and rescale
 the distribution function
 for want of the values of the KPZ parameters,
 in which case the difference
% among the GUE and GOE largest eigenvalue distributions
% and the Gaussian distribution
 among the predicted distributions and the Gaussian one is unpronounced.
They also had to discard data subject to
% ``avalanche'' events, i.e.,
 intermittent advance of burning fronts
 due to quenched disorder \cite{Miettinen_etal-EPJB2005}.
% of the paper \cite{Miettinen_etal-EPJB2005}.
Therefore, a quantitative test of Pr\"ahofer and Spohn's prediction
 has not been carried out so far.

\begin{figure}[t]
 \includegraphics[clip]{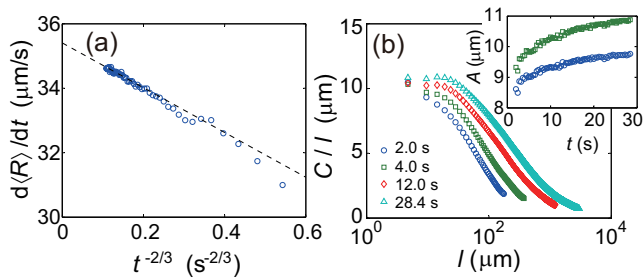}
 \caption{(Color online) Parameter estimation. (a) Growth rate $\rd \expct{R}/\rd t$ averaged over $1.0\unit{s}$ against $t^{-2/3}$. The $y$-intercept of the linear regression (dashed line) provides an estimate of $\lambda$. (b) $C(l,t)/l$ against $l$ for different times $t$. Inset: nominal estimates of $A$ obtained from $w(l,t)$ (blue bottom symbols) and $C(l,t)$ (green top symbols) as functions of $t$ (see text).\vspace{-5pt}}
 \label{fig3}
\end{figure}%

We first measure the value of $\lambda$
 experimentally.
For the circular interfaces,
 $\lambda$ is given as the asymptotic radial growth rate,
 which has a leading correction term as
% in the following form \cite{Krug_etal-Amar_Family-PRA1992}
%\begin{equation}
 $\lambda \simeq \rd\expct{R}/\rd t + a_v t^{-2/3}$
for $t \to \infty$ \cite{Krug_etal-Amar_Family-PRA1992}.
%  \label{eq:LambdaEstimation}
%\end{equation}
This relation is indeed confirmed in Fig.\ \ref{fig3}(a)
%The radial growth rate $\rd\expct{R}/\rd t$ is therefore
% plotted against $t^{-2/3}$ in Fig.\ \ref{fig3}(a).
%The linear relation between them confirms Eq.\ \eqref{eq:LambdaEstimation}
 and yields a precise estimate
% of $\lambda$
 at $\lambda = 35.40(23) \unit{\mu{}m/s}$.
% insensitive to the fitting range (inset).

The parameter $A$ can be determined, at least for flat interfaces,
% from the scaling of $C(l,t)$ and $w(l,t)$ 
 from the amplitude of $C(l,t)$ and $w(l,t)$ 
 through $C \simeq Al$ and $w^2 \simeq Al/6$
 in the limit $t \to \infty$
 \cite{Krug_etal-Amar_Family-PRA1992}.
Figure \ref{fig3}(b) shows $C(l,t)/l$ against $l$ for different times $t$.
% which would indicate the value of $A$ in the scaling regime for small $l$.
A similar series of plots is obtained for $6w^2/l$.
The value of $A$ can be estimated
 from the plateau level or the local maximum of these plots,
 but we find that these estimates
 increase slowly with time
 and do not agree with each other (inset).
This allows us to have only a rough estimate
% of A around
 $A \approx 10\unit{\mu{}m}$ for the range of time we study.
%This discrepancy and
%% logarithmic behavior, which implies
% the logarithmic correction to the Family-Vicsek scaling
%% for $l \ll l_*$
%% in the droplet geormetry,
% are yet to be explained.
%% theoretically.
%%A logarithmic behavior in the correlation, though for $l \gg l_*$,
%% was also shown in Ref.\ \cite{Escudero-PRL2008}.

\begin{figure}[t]
 \includegraphics[width=0.95\hsize,clip]{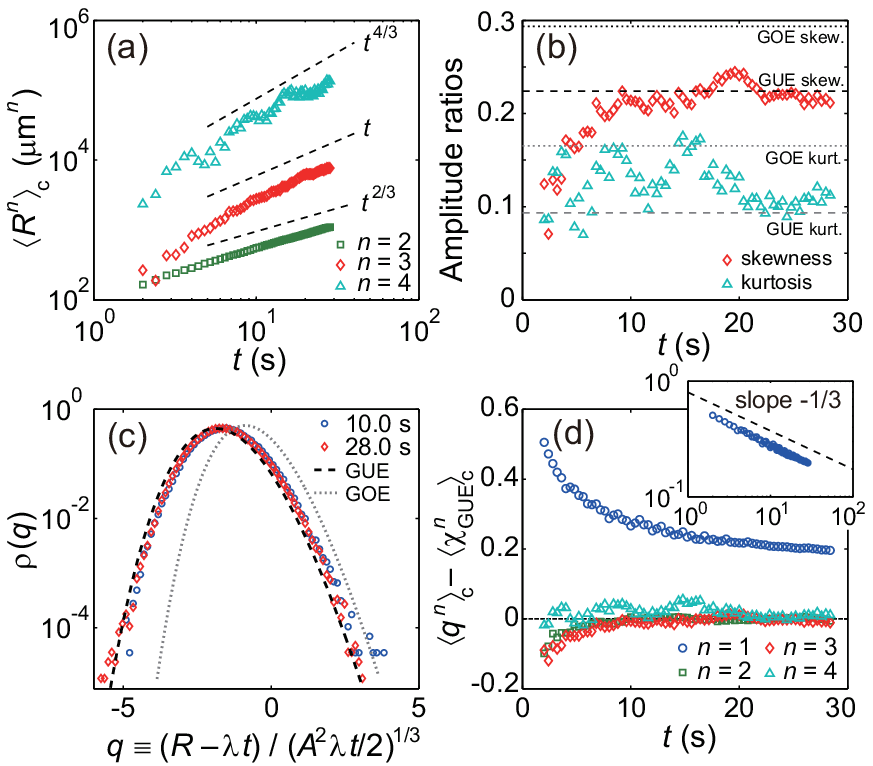}
 \caption{(Color online) Local radius distributions. (a) Cumulants $\expct{R^n}_{\rm c}$ vs $t$. The dashed lines are guides for the eyes showing the indicated powers. (b) Skewness $\expct{R^3}_{\rm c} / \expct{R^2}_{\rm c}^{3/2}$ and kurtosis $\expct{R^4}_{\rm c} / \expct{R^2}_{\rm c}^{2}$. The dashed and dotted lines indicate the values of the skewness and the kurtosis of the GUE and GOE TW distributions. (c) Local radius distributions as functions of $q \equiv (R-\lambda t) / (A^2\lambda t/2)^{1/3}$. The dashed and dotted lines show the GUE and GOE TW distributions, respectively. (d) Differences in the cumulants of $q$ and $\chi_{\rm GUE}$. The dashed line indicates $\expct{q^n}_{\rm c} = \expct{\chi_\mathrm{GUE}^n}_{\rm c}$. Inset: the same data for $n=1$ in logarithmic scales. The dashed line is a guide for the eyes.\vspace{-5pt}}
 \label{fig4}
\end{figure}%

%In any case, with these estimates of $\lambda$ and $A$,
Now we test Pr\"ahofer and Spohn's prediction
 for the circular interfaces:
%, it reads
\begin{equation}
 R(t) \simeq \lambda t + (A^2\lambda t/2)^{1/3} \chi_\mathrm{GUE}  \label{eq:PrahoferSpohn2}
\end{equation}
% with the largest eigenvalue $\chi_\mathrm{GUE}$ of GUE random matrices,
 with a random variable $\chi_\mathrm{GUE}$
% obeying the largest eigenvalue distribution
% of infinite-size GUE random matrices,
% or the GUE Tracy-Widom (TW) distribution \cite{Tracy_Widom-CommunMathPhys}.
 obeying the GUE TW distribution.
%A common quantity to characterize the distributions is
We first compute the cumulant
 $\expct{R^n}_{\rm c}$, for which Eq.\ \eqref{eq:PrahoferSpohn2} implies
 $\expct{R^n}_{\rm c} \simeq (A^2\lambda/2)^{n/3} \expct{\chi_\mathrm{GUE}^n}_{\rm c} t^{n/3}$ for $n \geq 2$.
Our data indeed show this power-law behavior in time [Fig.\ \ref{fig4}(a)],
 though higher order cumulants are statistically more demanding
 and hence provide less conclusive results.
We then calculate
 the skewness $\expct{R^3}_{\rm c} / \expct{R^2}_{\rm c}^{3/2}$
 and the kurtosis $\expct{R^4}_{\rm c} / \expct{R^2}_{\rm c}^{2}$,
 which do not depend on the parameter estimates.
The result in Fig.\ \ref{fig4}(b) shows that
 both amplitude ratios asymptotically converge to the values
 of the GUE TW distribution, about $0.2241$ for the skewness
 and $0.09345$ for the kurtosis \cite{Prahofer_Spohn-2000},
 and clearly rules out the GOE TW and Gaussian distributions.
Conversely, if we admit the GUE TW distribution,
 the amplitude of $\expct{R^2}_{\rm c}$ offers
 a precise estimate of $A$ at $9.98(7) \unit{\mu{}m}$,
 which is consistent with the direct estimate obtained above
 and hence used in the following.

%We then directly compare the experimentally measured distributions
% with the theoretical prediction of Eq.\ \eqref{eq:PrahoferSpohn2}.
Histograms of the local radius $R(x,t)$
 are then made and shown in Fig.\ \ref{fig4}(c)
 for two different times as functions
 of $q \equiv (R-\lambda t) / (A^2\lambda t/2)^{1/3}$,
 which corresponds to $\chi_{\rm GUE}$
 if Eq.\ \eqref{eq:PrahoferSpohn2} holds.
The experimental distributions show remarkable agreement
 with the GUE TW one without any fitting,
 apart from a slight horizontal translation.
% and are significantly different from the GOE TW and Gaussian distributions.
Indeed, time series of the difference between the $n$th order cumulants
 of $q$ and $\chi_{\rm GUE}$ [Fig.\ \ref{fig4}(d)] reveal that
 the second to fourth order cumulants of $q$ converge quickly
 to the GUE TW values, while the first order one, i.e., the mean,
 algebraically approaches it with a power close to $-1/3$ (inset).
This is theoretically reasonable behavior
 which stems from the existence of an additional constant term
 in Eq.\ \eqref{eq:PrahoferSpohn2}.
Therefore, we conclude that the local radii of the DSM2 nuclei
 asymptotically obey the GUE TW distribution
 at least up to the fourth order cumulants,
 confirming Pr\"ahofer and Spohn's prediction.

\begin{figure}[t]
 \includegraphics[width=0.7\hsize,clip]{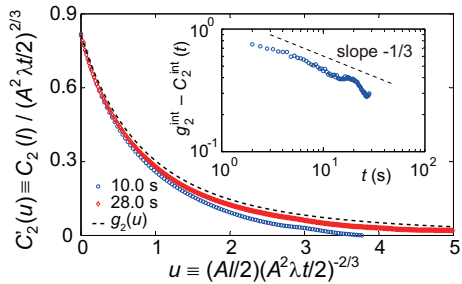}
 \caption{(Color online) Two-point correlation function $C_2(l,t)$ plotted in the rescaled axes $u \equiv (Al/2) (A^2\lambda t/2)^{-2/3}$ and $C'_2(u) \equiv C_2(l) / (A^2\lambda t/2)^{2/3}$. The dashed line indicates the theoretical prediction $g_2(u)$. Inset: integral of the rescaled correlation function $C_2^\mathrm{int} \equiv \int_0^\infty C'_2(u) \rd u$ as a function of time $t$. The difference from the theoretical counterpart $g_2^\mathrm{int} \equiv \int_0^\infty g_2(u) \rd u$ is shown. The dashed line is a guide for the eyes.\vspace{-5pt}}
 \label{fig5}
\end{figure}

We also measure the two-point correlation function
 $C_2(l,t) \equiv \expct{R(x+l,t)R(x,t)} - \expct{R}^2$.
Theory predicts that $C_2(l,t)$ is asymptotically described
 by the Airy$_2$ process $\mathcal{A}_2(t)$
 or by the dynamics of the largest eigenvalue in Dyson's Brownian motion
% with $\beta=2$
 of GUE matrices \cite{Mehta-Book2004}
 as $C_2(l,t) \simeq (A^2\lambda t/2)^{2/3} g_2(u)$
 with $g_2(u) \equiv \expct{\mathcal{A}_2(u+t)\mathcal{A}_2(t)}$
 and $u \equiv (Al/2) (A^2\lambda t/2)^{-2/3}$
 \cite{Prahofer_Spohn-JStatPhys2002}.
Our experimental data confirm this
 with an algebraic finite-time correction consistent with the power $-1/3$
 (Fig.\ \ref{fig5}).

In comparison with past experimental studies showing diverse scalings,
% in the interface roughening,
 one may wonder why the liquid crystal turbulence
 exhibits such clear KPZ-class behavior.
We consider that the following three factors are essential:
(a) The growth of DSM2 results from strictly local processes
 due to the turbulent flow on the interfaces
%There is no contribution
 and not from inward or outward interactions of the cluster,
 which could induce long-range effects and affect the universality.
(b) The stochasticity of the process stems
 from intrinsic turbulent fluctuations
% and does not rely on any inhomogeneity of the cell, which would cause
 overwhelming quenched disorder.
(c) Good controllability and fast response of the liquid crystals
 allowed us to repeat hundreds of experiments in the same conditions,
 leading to statistically reliable data.
The reproducibility of the presented results was confirmed
 with different voltages and spatial resolutions
 with the same quality of data (not shown).

In conclusion, measuring the growth of DSM2 nuclei
 in the electroconvection,
 we have found the circular interface roughening
 clearly characterized by the scaling laws of the KPZ class
 in 1+1 dimensions.
%We have also revealed that 
%The amplitude of the Family-Vicsek scaling is,
% however, found to grow logarithmically in time for $l \ll l_*$
% as opposed to flat KPZ interfaces
% \cite{Krug_etal-Amar_Family-PRA1992}.
%The characteristic exponents are the same as for flat KPZ interfaces,
%The amplitude of the Family-Vicsek scaling
% is, however, found to grow logarithmically in time for $l \ll l_*$
% as opposed to flat KPZ interfaces
% \cite{Krug_etal-Amar_Family-PRA1992}.
Moreover, we have shown
% without any fitting parameters
 without fitting
 that the fluctuations of the cluster local radius asymptotically obey
 the Tracy-Widom distribution of the GUE random matrices
 and revealed the finite-time effect.
Together with the agreement in the two-point correlation, 
 our experimental results quantitatively confirm 
% for the first time in experiments
 the geometry-dependent universality of the (1+1)-dimensional KPZ class
 prescribing detailed information of the scale-invariant fluctuations.
In this respect, investigations of flat interfaces in the same system
 are of outstanding importance and are in progress.

We acknowledge enlightening discussions
 with H. Chat\'e, M. Pr\"ahofer, T. Sasamoto, and H. Spohn.
We also thank M. Pr\"ahofer and F. Bornemann for providing us
 with numerical values of the TW distributions and the covariance
 of the Airy$_2$ process.
This work is partly supported by JSPS and by MEXT (No. 18068005).

\textit{Note added in proof.}
After submission of this Letter,
 Sasamoto and Spohn reported an exact solution
 of the (1+1)-dimensional KPZ equation \cite{Sasamoto_Spohn-PRL},
 which offers a clear theoretical ground of our experimental results.

\end{document}